\documentclass[twocolumn,pra,showpacs,superscriptaddress,showpacs]{revtex4-1}
\usepackage{graphicx}
\usepackage{amssymb,amsmath}
\usepackage{bm}
\usepackage{dcolumn}
\usepackage{subfigure}
\usepackage{float}
\usepackage{url}
\usepackage{color}
\usepackage{txfonts}
\usepackage[driverfallback=dvipdfm]{hyperref}
\hypersetup{pdfpagemode=FullScreen,colorlinks=true,breaklinks,urlcolor=blue,linkcolor=blue,citecolor=blue}

\begin{document}
\title{Measuring Hopf links and Hopf invariants in a quenched topological Raman lattice}

\author{Jinlong Yu}
\affiliation{Institute for Advanced Study, Tsinghua University, Beijing 100084, China}

\begin{abstract}
In a recent experimental work [Z. Wu \emph{et al.}, Science {\bf{354}}, 83 (2016)], the PKU-USTC group realized a two-dimensional two-band quantum anomalous Hall model on a square Raman lattice. By quenching the atom-laser detuning of such a Raman lattice, the time-dependent Bloch vectors for each quasimomentum points define a Hopf mapping from quasimomentum-time space $(k_x,k_y,t)\in T^3$ to the Bloch sphere $S^2$. The Hopf links between the preimages of any two Bloch vectors on $S^2$ can be measured experimentally through Bloch state tomography using spin-resolved time-of-flight measurement together with suitable radio-frequency manipulations. The dynamical Hopf invariants, which are quasimomentum-time integrations of Chern-Simons densities, can also be extracted experimentally through certain quench processes. As the Hopf invariant equals the Chern number of the post-quench Hamiltonian, the measurements of Hopf links and Hopf invariants provide an alternative way of understanding the topological phase diagram of the equilibrium system. 
\end{abstract}

\maketitle

\section{Introduction}
Ultracold atoms in lattice systems are powerful simulators for studying topological physics~\cite{Bloch2008,Dalibard2011,Nori2014,Goldman2014b,Cooper2018}. A case in point is the Harper-Hofstadter model~\cite{Harper1955,Hofstadter1976}, wherein the ground state of Hamiltonian is characterized by a topological invariant -- the Chern number -- has been studied experimentally~\cite{Aidelsburger2013, Miyake2013, Aidelsburger2015, Ketterle2015} using the laser-assisted-tunneling technique. In another seminal experiment, the Haldane model~\cite{Haldane1988} on a hexagonal lattice is realized by the ETH group~\cite{Jotzu2014} via the shaking-lattice technique~\cite{Zheng2014}. Along this line, a Haldane-like model was later achieved by the Hamburg group~\cite{Sengstock2016a,Sengstock2018,Sengstock2017}. 
Recently, the PKU-USTC group has implemented the quantum anomalous Hall (QAH) model on a square lattice~\cite{USTC_SOC2016}. Key to their setup is a unconventional (spin-dependent) Raman lattice with coupled hyperfine spin states, aside from the conventional (spin-independent) optical lattices.

Cold-atom systems also present ideal platform for studying the out-of-equilibrium physics, in particular, the quench dynamics~\cite{Caio-Cooper2015, Caio-Cooper2016, Hu-Zoller2016, Mueller2016, Wilson2016, Wang2017,Yu2017,Sengstock2018,Sengstock2017}. In Refs.~\cite{Wang2017,Yu2017}, the quench dynamics of a generic two-dimensional two-band model initialized in a topologically trivial state is theoretically explored. Remarkably, such $(2+1)$-dimensional quenched system naturally allows for a Hopf mapping from the quasimomentum-time space $(k_x,k_y,t)\in T^3$ to the Bloch sphere $S^2$. Subsequent experiments have measured, based on a quenched Haldane-like model, both trivial and nontrivial Hopf links~\cite{Sengstock2018,Sengstock2017} between the preimages of Bloch vectors pointing to the north and south poles on $S^2$, which is achieved through the quasimomentum-resolved Bloch-state tomography for the azimuthal phase~\cite{Hauke_Lewenstein2014,Sengstock2016a}. 

However, the Bloch-state tomography as in Refs.~\cite{Hauke_Lewenstein2014,Sengstock2016a} is not complete, in the sense that the sign of the $z$ component of the Bloch vectors can not be resolved directly. By contrast, the topological Raman lattice system~\cite{USTC_SOC2016} allows for a complete Bloch-state tomography, which is possible through a combination of the spin-resolved time-of-flight measurement~\cite{USTC_SOC2016,Liu2013} and suitable radio-frequency (rf) manipulations rotating the spin basis~\cite{Monroe2017}. This paves the way to a direct visualization of the Hopf links between the preimages of any two Bloch vectors, and moreover, an extraction of the topological number hidden in the quench dynamics, i.e., the Hopf invariant. We note that, although there is a recent experimental measurement of the Hopf invariant in a single nitrogen-vacancy center~\cite{Duan2017}, the corresponding measurement in a \emph{lattice} system is still to be realized. Our current protocol fulfills this task in the theoretical side, and it is likely to be realized experimentally in the near future.

This paper is organized as follows. In Sec.~\ref{Sec:Hopf_Links}, we describe the scheme to measure the Hopf links in a topological Raman lattice. In particular, we will show that the QAH model is adopted to describe the topological Raman lattice. In the simplest setup, with spin-resolved time-of-flight measurement, nontrivial Hopf links can be obtained from the trajectories with maximal spin imbalance for the quenched QAH model. Additional rf manipulations together with the spin-resolved time-of-flight measurement can fulfill the Bloch state tomography, where Hopf links for more than two preimages can be visualized. In Sec.~\ref{Sec:Measure_Hopf_inv}, we define the dynamical Hopf invariant for the quenched QAH model and deduce its relation with the Chern number of the post-quench Hamiltonian. We then show how the Hopf invariant can be evaluated from the experimentally measured Bloch vectors.  We conclude in Sec.~\ref{Sec:Conclusion}. 

\section{Hopf links in a topological Raman lattice}\label{Sec:Hopf_Links}
\subsection{The quantum anomalous Hall model for the topological Raman lattice}\label{Sec:Link_QAH}
In the experiment by the PKU-USTC group~\cite{USTC_SOC2016}, the following two-dimensional two-band QAH Hamiltonian is realized on a square Raman lattice (length is measured in units of the lattice constant):
\begin{equation}\label{Eq:Hami_QAH}
  H\left( {\mathbf{k}} \right) = {\mathbf{h}}\left( {\mathbf{k}} \right) \cdot {\bm{\sigma }},
 \end{equation}
 where 
 \begin{equation}
 \begin{aligned}
  &h_x =  J_{\text{so}}{\sin {k_x}}, \quad h_y =  J_{\text{so}}{\sin {k_y}} , \\
   &h_z =   {M - J_0(\cos {k_x} - \cos {k_y})}.
  \end{aligned}
\end{equation}
The space of the Pauli matrices  $\bm{\sigma} = (\sigma _x, \sigma _y,\sigma _z)$ are spanned by the two hyperfine spin states, which are coupled by Raman lasers. Here ${\bf k}=(k_x, k_y)$ is the quasimomentum, $J_0$ and $J_{\text{so}}$ 
are respectively the spin-conserved and spin-flip hopping coefficients,  and $M$ arises from the detuning between the Zeeman splitting of the two spin states and the Raman lasers. For simplicity, we take $J_0 = J_{\text{so}} = 1$ as the energy unit. In typical experiments, they can take different values, and are of the order of a few hundred hertz~\cite{USTC_SOC2016}. The spectrum of the system is $E_\pm=\pm h$, where $h=|\bf{h}| = \sqrt{h_x^2 + h_y^2+h_z^2}$, which can be measured experimentally~\footnote{Generally, there is a $h_0 I_2$ term in the Hamiltonian, where $I_2$ is a $2\times2$ identity matrix. Such a term does not affect the topology of the system, but shifts the energies of the two bands to be $E_\pm = h_0 \pm h$. Thus we get $h = (E_+ - E_-)/2$ from the experimentally measured spectra $E_\pm$.} using the spin-injection spectroscopy technique~\cite{Wang-Zhang2012,Cheuk-Zwierlein2012}. 

\begin{figure}[t]
  \includegraphics[width=\columnwidth]{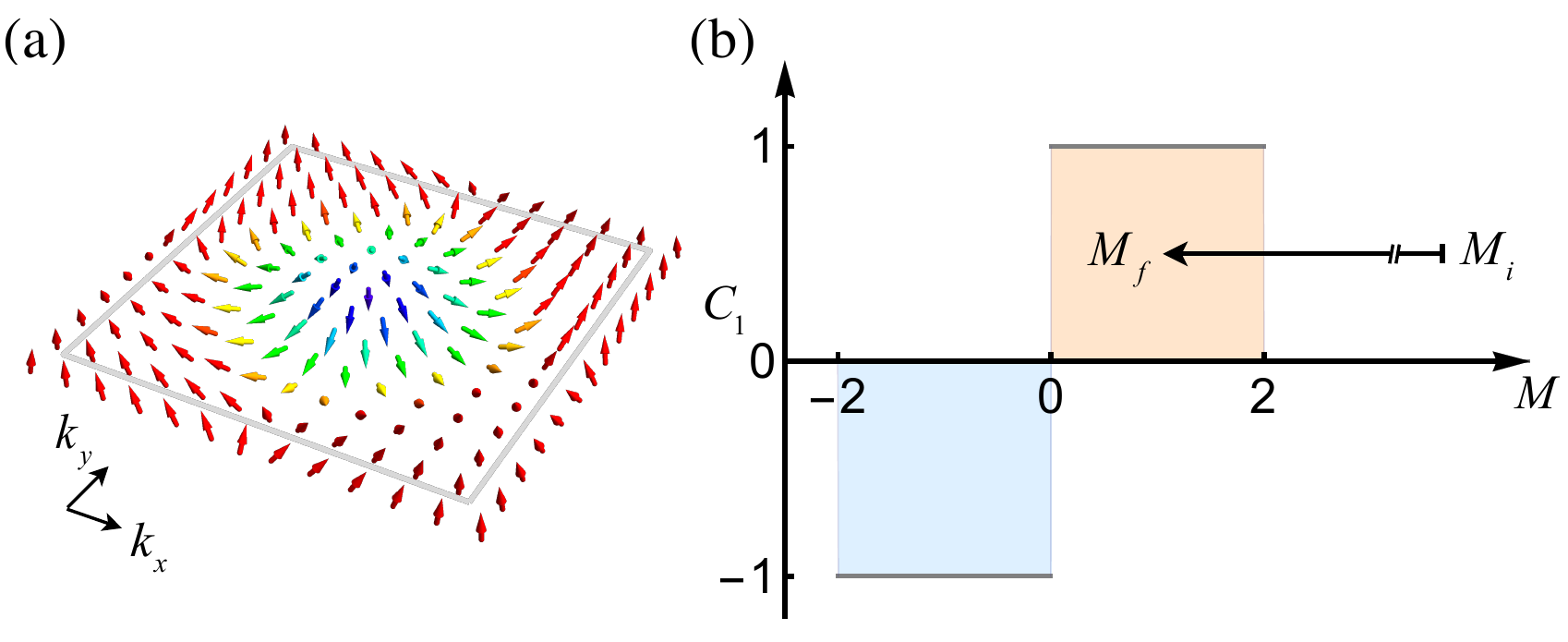}\\
  \caption{A schematic illustration for the quenched QAH model. (a) The normalized Hamiltonian vector ${ \mathbf{\hat h}}(\mathbf{k}) = \mathbf{h}/|\mathbf{h}|$ in the quasimomentum space for the topologically nontrivial case with $M=0.6$. The gray square shows the first Brillouin zone. (b) The topological phase diagram of the QAH model and the quench protocol. The detuning $M$ is quenched from $M_i$ to $M_f$. The state of the system is initially prepared as the ground state of the Hamiltonian with $M=M_i$, then evolves under the Hamiltonian with $M=M_f$. We consider the case that the initial state is topologically trivial ($|M_i|>2$), and the final Hamiltonian can be both trivial or nontrivial.}\label{Fig1}
\end{figure}

The topology of the Hamiltonian can be characterized by the (first) Chern number~\cite{TKNN1982} of the lowest band~\cite{Qi2006}, i.e., 
\begin{equation}\label{Eq:C_1_Qi}
  {C_1} = \int_{{\text{BZ}}} {\frac{{{d}k_x}{d{k_y}}}{{4\pi }}}{\bf{\hat h}} \cdot \left( {{\partial _{{k_x}}}{\bf{\hat h}} \times {\partial _{{k_y}}}{\bf{\hat h}}} \right),
\end{equation}
with ${\mathbf{\hat h}} = {\mathbf{h}}/|\mathbf{h}|$. The integration is performed in the first Brillouin zone $({k_x},{k_y}) \in [ - \pi ,\pi ) \times [ - \pi ,\pi )$. There, the winding patten of the vector ${\mathbf{\hat h}}$ [see, e.g., Fig.~\ref{Fig1}(a)] as a function of detuning $M$ gives rise to
\begin{equation}
  {C_1}\left( M \right) = \left\{ \begin{gathered}
  1, \hfill \\
   - 1, \hfill \\
  0, \hfill \\
\end{gathered}  \right.\quad \begin{array}{*{20}{c}}
  {0 < M < 2;} \\
  { - 2 < M < 0;} \\
  {{\text{other cases}}{\text{,}}}
\end{array}
\end{equation}
which determines the topological phase diagram of the system [Fig.~\ref{Fig1}(b)].

\subsection{Quench dynamics for the QAH model}
We now consider the quench dynamics for the QAH model [Fig.~\ref{Fig1}(b)], following Ref.~\cite{Yu2017}. We consider the case that the detuning $M$ is time dependent, i.e., 
\begin{equation}
  M(t) = \left\{ \begin{gathered}
  {M_i},\quad t \leqslant 0; \hfill \\
  {M_f},\quad t > 0. \hfill \\
\end{gathered}  \right.
\end{equation}
We use capital Greek letters ${\Theta _{\mathbf{k}}}$ and ${\Phi_{\mathbf{k}}}$ to parameterize the Hamiltonian vector ${\mathbf{\hat h}}\left( {\mathbf{k}} \right) = {\bf{h}}\left( {\mathbf{k}} \right)/|{\bf{h}}\left( {\mathbf{k}} \right)|$, with ${\mathbf{\hat h}}\left( {\mathbf{k}} \right) = (\sin {\Theta _{\mathbf{k}}}\cos {\Phi _{\mathbf{k}}},\sin {\Theta _{\mathbf{k}}}\sin {\Phi _{\mathbf{k}}},\cos {\Theta _{\mathbf{k}}})$. We prepare the initial state as the ground state of the Hamiltonian $H_i= {\bf{h}}_i\cdot\bm{\sigma}$, or written explicitly,
\begin{equation}
\left| {{\psi _{\mathbf{k}}}\left( {t = 0} \right)} \right\rangle  = \left( \begin{gathered}
  \sin \frac{{\Theta _{\mathbf{k}}^i}}{2} \hfill \\
   - \cos \frac{{\Theta _{\mathbf{k}}^i}}{2}{e^{i\Phi _{\mathbf{k}}^i}} \hfill \\
\end{gathered}  \right).
\end{equation}

After the quench, the state evolves under the new Hamiltonian $H_f= {\bf{h}}_f\cdot\bm{\sigma}$, as described by $\left| {{\psi _{\mathbf{k}}}\left( t \right)} \right\rangle  = {e^{ - i{H_f}t}}\left| {{\psi _{\mathbf{k}}}\left( {t = 0} \right)} \right\rangle $. Note that while the quench changes the $h_z({\bf k})$ term through modulations of $M$, both $h_x({\bf k})$ and $h_y({\bf k})$ remain unaltered. Hence, the azimuthal phase ${\Phi_{\mathbf{k}}}$ stays invariant through the quench dynamics, i.e., ${\Phi _{\mathbf{k}}^i} = {\Phi_{\mathbf{k}}^f}$, so that we have
\begin{equation} \label{Eq:psi_k_t}
  \begin{aligned}
  \left| {{\psi _{\mathbf{k}}}\left( t \right)} \right\rangle  = &\cos \left( {{h_f}t} \right)\left| {{\psi _{\mathbf{k}}}\left( {t = 0} \right)} \right\rangle  \hfill \\
  &  + i\sin \left( {{h_f}t} \right)\left( \begin{gathered}
  \sin \left( {\Theta _{\mathbf{k}}^f - \frac{{\Theta _{\mathbf{k}}^i}}{2}} \right) \hfill \\
   - \cos \left( {\Theta _{\mathbf{k}}^f - \frac{{\Theta _{\mathbf{k}}^i}}{2}} \right){e^{i\Phi _{\mathbf{k}}^i}} \hfill \\
\end{gathered}  \right), \hfill \\
\end{aligned}
\end{equation}
where $h_f = |{\bf{h}}_f|$. Parameterizing the state (\ref{Eq:psi_k_t}) on a Bloch sphere as
\begin{equation}\label{Eq:psi_theta_phi}
  \left| {{\psi _{\mathbf{k}}}\left( t \right)} \right\rangle \equiv  \left( \begin{gathered}
  \sin \left( {{\theta _{\mathbf{k}}}/2} \right) \hfill \\
   - \cos \left( {{\theta _{\mathbf{k}}}/2} \right){e^{i{\phi _{\mathbf{k}}}}} \hfill \\
\end{gathered}  \right),
\end{equation}
the corresponding time- and quasimomentum-dependent Bloch vectors are given by
\begin{equation} \label{Eq:n_vector}
  \begin{aligned}
  {{\mathbf{n}}}\left({\mathbf{k}}, t \right) &= - \left\langle {{\psi _{\mathbf{k}}}\left( t \right)} \right|{\bm{\sigma }}\left| {{\psi _{\mathbf{k}}}\left( t \right)} \right\rangle  \hfill \\
   &= \left( {\sin {\theta _{\mathbf{k}}}\cos {\phi _{\mathbf{k}}},\sin {\theta _{\mathbf{k}}}\sin {\phi _{\mathbf{k}}},\cos {\theta _{\mathbf{k}}}} \right). \hfill \\
\end{aligned}
\end{equation}
Since the $(k_x, k_y, t)$ represents a point on the $3$-torus ($T^3$), the set of 
Eqs.~(\ref{Eq:psi_k_t})-(\ref{Eq:n_vector}) define a mapping from $T^3$ to the Bloch sphere $S^2$, also known as the Hopf mapping~\cite{Moore-Wen2008} (we will discuss this point in more details in Sec.~\ref{Sec:Measure_Hopf_inv}). For a Bloch vector ${{\mathbf{n}}}$, i.e., a point on $S^2$ as parameterized by $({\theta} , {\phi })$, the collection of $(k_x, k_y, t)$ points mapping to this $({\theta }, {\phi })$ form a loop, which represents a \emph{preimage}  of ${{\mathbf{n}}}$.  The preimages of two different Bloch vectors $\bf{n}$ and $\bf{n}'$ can be linked, and the linking number equals to the Chern number of the post-quench Hamiltonian $H_f$~\cite{Wang2017}.

\subsection{Measuring the linking number from the trajectories with maximal spin imbalance}

\begin{figure*}[tp]
  \includegraphics[width=\textwidth]{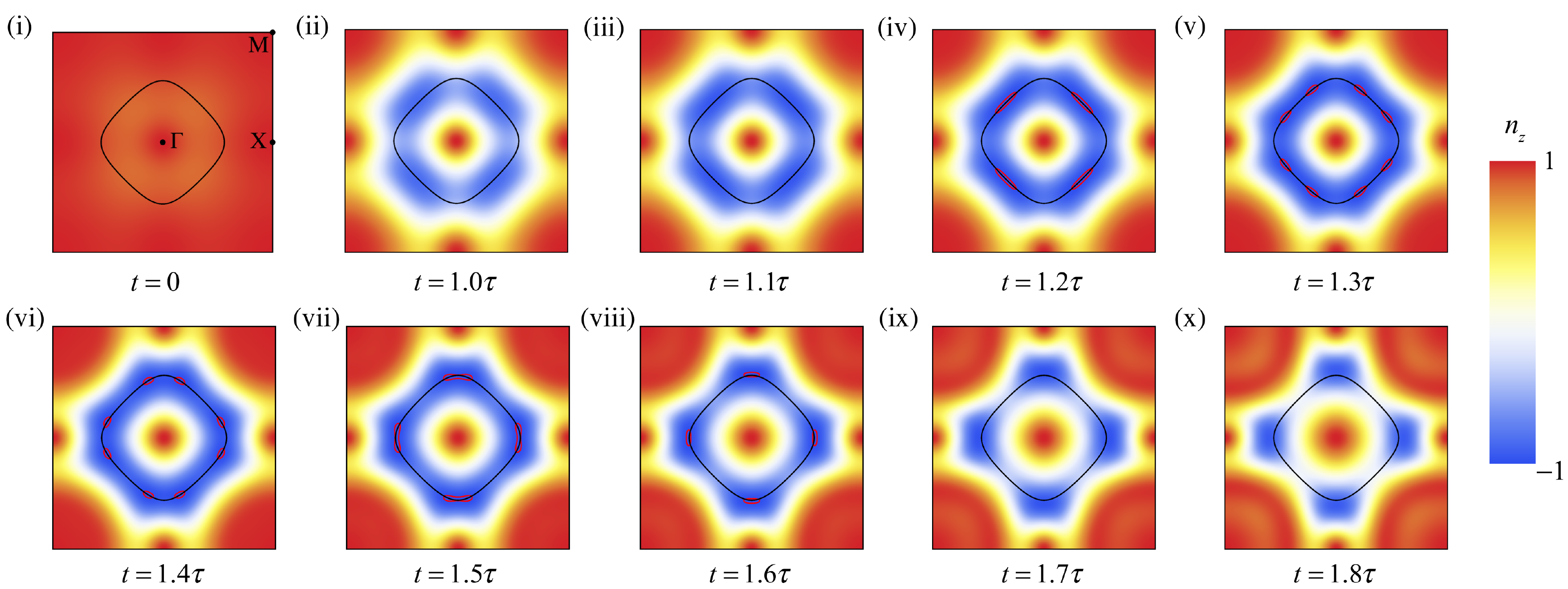}\\
  \caption{The quasimomentum resolved, time-dependent spin imbalance $n_z({\bf{k}},t)$ of the quenched QAH model.  The system is initially prepared as the eigenstate of the Hamiltonian with $M_i = 3$ (i); it then evolves under the Hamiltonian with $M_f = 0.6$ (ii-x). High symmetric points in the first Brillouin zone are marked in (i), where the corresponding $(k_x, k_y)$ coordinates are $\Gamma (0, 0)$, X$(\pi,0)$, and M$(\pi,\pi)$.  During the time evolution, $n_z = 1$ is maintained for these three points. 
  The loop around the $\Gamma$ point is determined by Eq.~(\ref{Eq:south-pole_k}) (with $M_i = 3$ and $M_f = 0.6$), which indicates the quasimomentum points that can reach $n_z = -1$. The additional red contours in (iv-viii) indicate the quasimomentum points with $n_z = -0.99$.
  The time unit $\tau = 1/J_0$ is about several miniseconds for a typical experiment~\cite{USTC_SOC2016}.}\label{Fig2}
\end{figure*}

\begin{figure}[tbp]
  \includegraphics[width=\columnwidth]{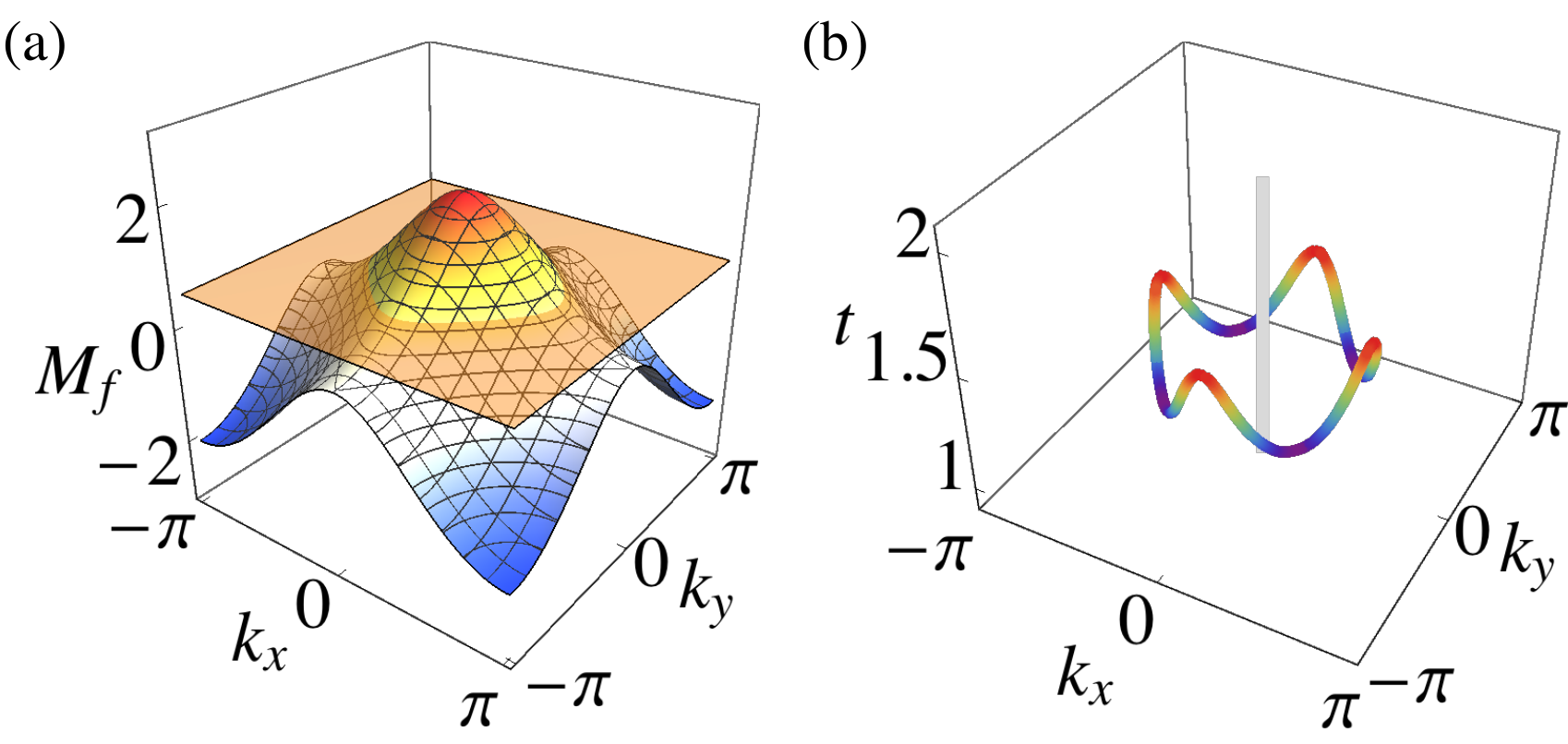}\\
  \caption{Extracting the Hopf link from preimages with maximal spin imbalance. (a) The surface satisfies Eq.~(\ref{Eq:south-pole_k}) for different $M_f$ values, with $M_i = 3$. As $M_f$ varies, it shrinks to the $\Gamma$ point for $M_f = 2$, and to the M point for  $M_f = -2$. The intersection between this surface and the constant plane $M_f=0.6$ gives rise to the contours shown in Fig.~\ref{Fig2}. (b) The trajectories with maximal spin imbalance in the $(k_x,k_y,t)$ space: straight line for $n_z=1$; loop for $n_z = -1$ [determined by Eqs.~(\ref{Eq:south-pole_k}) and (\ref{Eq:south-pole_t})]. The linking number between them (i.e., preimages of $n_z=\pm1$) equals the Chern number of the final Hamiltonian $C_1=1$ for this case. Such a Hopf link has been measured experimentally by the PKU-USTC group~\cite{USTC_2018,LinkingLiu2018}.}\label{Fig3}
\end{figure}

In Refs.~\cite{Sengstock2018,Sengstock2017}, the phase vortices - the vortex pattern of the azimuthal phase $\phi_{\mathbf{k}}$ in the quasimomentum space -  of a \textit{quenched} Haldane-like model are measured experimentally. As is found, two types of phase vortices can appear: one is static, which corresponds to the Bloch vector pointing to the north pole of the Bloch sphere ($n_z = 1$); the other is dynamical, which corresponds to the Bloch vectors pointing to the south pole ($n_z=-1$). The preimages for the north and south poles (i.e., trajectories for the static and dynamical vortices) are linked if the final Hamiltonian is topological, as opposed to the case with trivial final Hamiltonian where nontrivial linking is absent~\cite{Yu2017,Wang2017,Sengstock2018,Sengstock2017}. 

For the QAH model~(\ref{Eq:Hami_QAH}) realized by the PKU-USTC group, the Pauli matrices represent two hyperfine spin states, rather than the sublattice degrees of freedom for a single spin state as in Refs.~\cite{Sengstock2018,Sengstock2017}. As such, a spin-resolved time-of-flight measurement~\cite{USTC_SOC2016} can be exploited to record the time- and quasimomentum-dependent spin imbalance $n_z({\bf{k}},t)$, thus allows for a direct measurement of the Hopf links between preimages of $n_z = 1$ and $n_z=-1$. Specifically, the parameter equations for $n_z = -1$ can be derived from Eq.~(\ref{Eq:psi_k_t}) as,
\begin{eqnarray}\label{Eq:south-pole_k}
  \cos \left( {\Theta _{\mathbf{k}}^f - \frac{{\Theta _{\mathbf{k}}^i}}{2}} \right) &&= 0, \\
  \label{Eq:south-pole_t}
\cos \left( {{h_f}t} \right) &&= 0.
\end{eqnarray}
A typical measurement of the Hopf link between $n_z = \pm1$ is shown in Figs.~\ref{Fig2} and \ref{Fig3}.

\subsection{pi/2-rf pulses and spin-resolved Bloch state tomography: mutual links for more than two preimages}
As described earlier, measurement of the spin imbalance renders extraction of the preimages for the $n_z = \pm 1$ points. Note that when the post-quench Hamiltonian is topologically trivial, the preimage for $n_z = -1$ vanishes and the linking is absent. 

One can proceed further to measure both the $n_x$ and $n_y$ components, by applying $\frac{\pi}{2}$-rf pulses in the $x$-$y$ plane before taking the Stern-Glarch measurement~\cite{Alba2011,Lisle2014,Monroe2017}. We note that, the rf pulses can rotate the spin basis as follows
\begin{equation}
\begin{gathered}
  {e^{i(\pi /2){\sigma _x}/2}}{\sigma _z}{e^{ - i(\pi /2){\sigma _x}/2}} = {\sigma _y}, \hfill \\
  {e^{i(\pi /2){\sigma _y}/2}}{\sigma _z}{e^{ - i(\pi /2){\sigma _y}/2}} =  - {\sigma _x}. \hfill \\ 
\end{gathered} 
\end{equation}
A $\frac{\pi}{2}$-rf pulse along the $x$ direction imprints a phase factor to the wavefunction as 
\begin{equation}
\left| {{{ \psi }_{\mathbf{k}}^\prime}\left( t \right)} \right\rangle  = {e^{ - i(\pi /2){\sigma _x}/2}}\left| {{\psi _{\mathbf{k}}}\left( t \right)} \right\rangle .
\end{equation}
A subsequent Stern-Glarch measurement then gives rise to 
\begin{equation}
\begin{aligned}
  \left\langle {{{ \psi }_{\mathbf{k}}^\prime}\left( t \right)} \right|{\sigma _z}\left| {{{ \psi }_{\mathbf{k}}^\prime}\left( t \right)} \right\rangle  &= \left\langle {{\psi _{\mathbf{k}}}\left( t \right)} \right|{e^{i(\pi /2){\sigma _x}/2}}{\sigma _z}{e^{ - i(\pi /2){\sigma _x}/2}}\left| {{\psi _{\mathbf{k}}}\left( t \right)} \right\rangle  \hfill \\
   &= \left\langle {{\psi _{\mathbf{k}}}\left( t \right)} \right|{\sigma _y}\left| {{\psi _{\mathbf{k}}}\left( t \right)} \right\rangle  =  - {n_y}\left( {{\mathbf{k}},t} \right). \hfill \\ 
\end{aligned} 
\end{equation}
Likewise, a $\frac{\pi}{2}$-rf pulse along the $y$ direction followed by the Stern-Glarch measurement gives the ${n_x}\left( {{\mathbf{k}},t} \right)$ component. 

\begin{figure}[tbp]
  \includegraphics[width=\columnwidth]{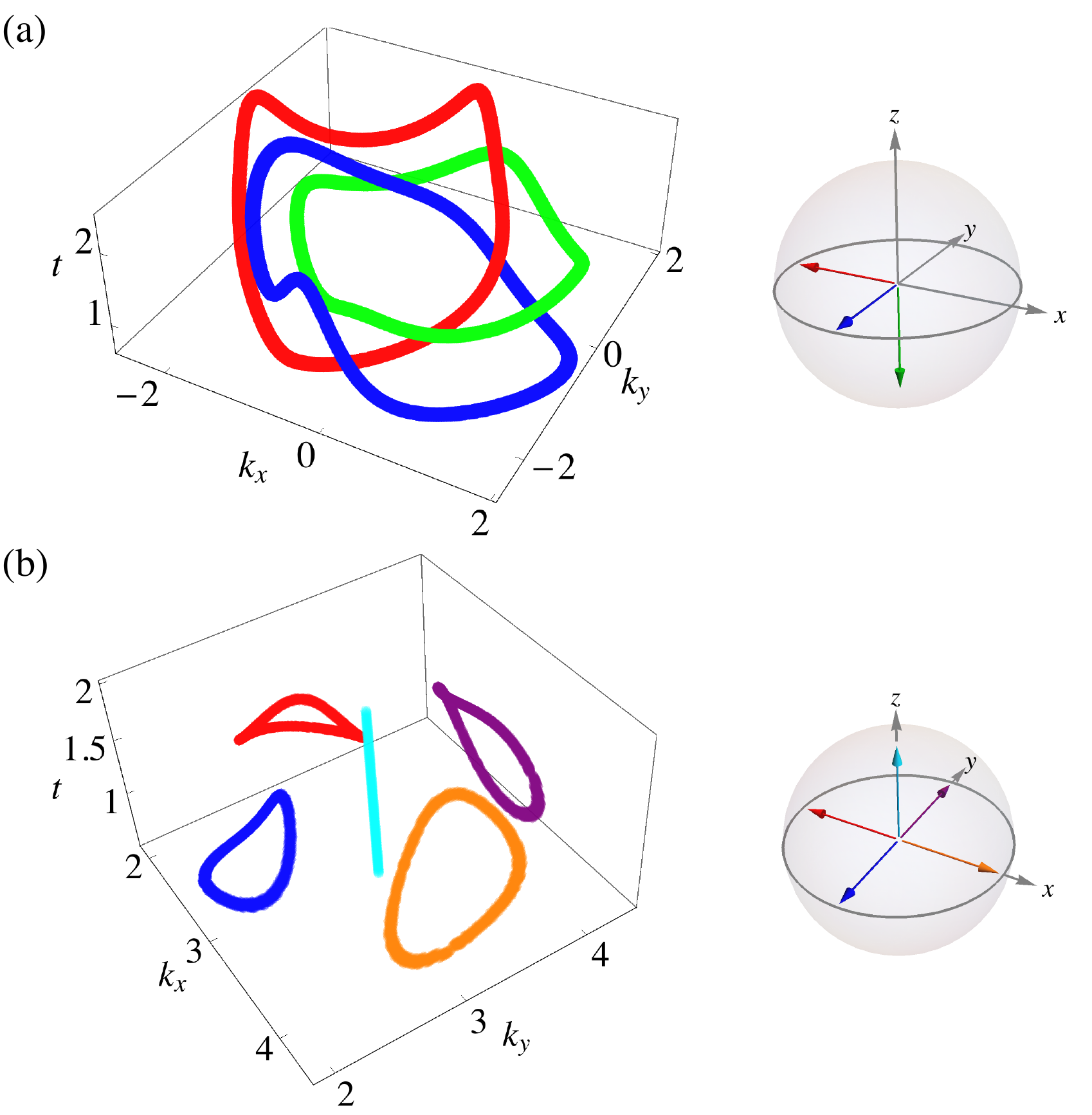}\\
  \caption{Hopf links for more than two preimages. (a) The nontrivial mutual linking between preimages of $n_x = -1$ and $n_y = -1$ and $n_z = -1$ points for the case of a topological quench ($M_i = 3 \to M_f = 0.6$, i.e., the same as Fig.~\ref{Fig2}). (b) The trivial linking between preimages of $n_x = \pm1$ and $n_y = \pm1$ and $n_z = 1$ points for a topologically trivial quench ($M_i = 3 \to M_f = -2.5$). The preimage for $n_z=-1$ vanishes for this case.}\label{Fig4}
\end{figure}

Once the $\left( {n_x\left( {{\mathbf{k}},t} \right), n_y\left( {{\mathbf{k}},t} \right), n_z\left( {{\mathbf{k}},t} \right) } \right)$ are measured, the Bloch state tomography is achieved~\footnote{Such an approach can be compared with the existing experiments by the Hamburg group~\cite{Sengstock2018, Sengstock2017}, where the sign of $n_z$ can not be determined directly.}. This enables the extraction of the preimages for any points on the Bloch sphere; see typical results for the Hopf links of certain preimages as summarized in Fig.~\ref{Fig4}.

\section{Measuring the Hopf invariant from the quench dynamics}\label{Sec:Measure_Hopf_inv}
In Ref.~\cite{Wang2017}, we have proved a theorem, which states that the dynamical Hopf invariant of a quenched two-band model initialized in a topologically trivial state equals the Chern number of the post-quench Hamiltonian. The proof presented therein makes use of a composite mapping. Below we will present an alternative yet more concise proof. Thereafter, we will show how the Hopf invariant, and thus the Chern number, can be extracted from the experimentally measured $\left( {n_x\left( {{\mathbf{k}},t} \right), n_y\left( {{\mathbf{k}},t} \right), n_z\left( {{\mathbf{k}},t} \right) } \right)$.

\subsection{Dynamical Hopf invariant and its relation to Chern number}

\begin{figure*}[htbp]
  \includegraphics[width=1.0\textwidth]{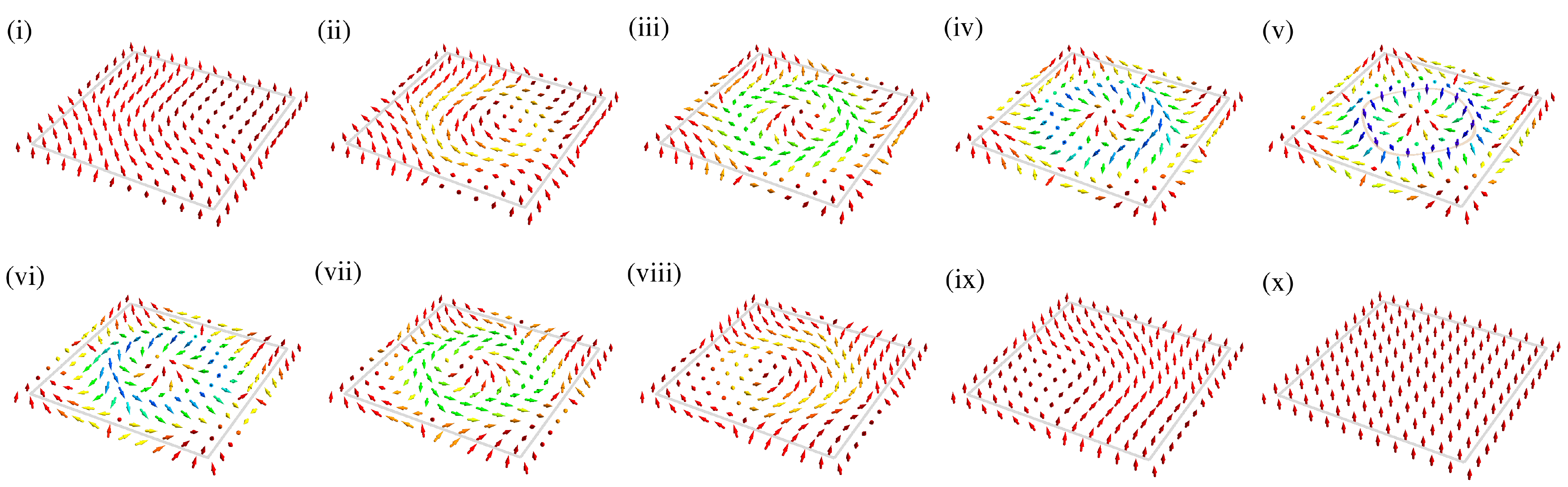}\\
  \caption{The quasimomentum resolved Bloch vectors [Eq.~(\ref{Eq:n_x_y_z})] of the quenched QAH model ($M_i = +\infty\to M_f = 0.6$), for different rescaled times: (i) ${\tilde t} = 0.2\pi$, (ii) ${\tilde t} = 0.4\pi$, ..., (x) ${\tilde t} = 2\pi$. The gray squares represent the first Brillouin zone. The configuration of the Bloch vectors returns back to its initial one (i.e., all the Bloch vectors pointing to the north pole direction) at time ${\tilde t} = 2\pi$, which shows the periodicity in the rescaled-time direction. There is an additional contour (determined by $\hat h_z = 0$) in (v) with ${\tilde t} = \pi$. The Bloch vectors located at this contour at ${\tilde t} = \pi$ all point to the south pole direction. }\label{Fig5}
\end{figure*}

To extract the topological invariants from the quench dynamics, the concrete value of $M_i$ is not important as long as we take $|M_i|>2$. Hence, for simplicity, we can take it to be infinity and set the initial state as the spin polarized state with all the Bloch vectors pointing to the north pole of the Bloch sphere:
\begin{equation}
  \left| {{\psi _{\mathbf{k}}}\left( {t = 0} \right)} \right\rangle  = \left( \begin{gathered}
  0 \hfill \\
  1 \hfill \\
\end{gathered}  \right).
\end{equation}
The time-evolved state under the Hamiltonian ${H}_f\left( {\mathbf{k}} \right) = {\mathbf{h}_f}\left( {\mathbf{k}} \right) \cdot {\bm{\sigma}}$ is described by 
\begin{equation}\label{Eq:psi_t}
  \begin{aligned}
  &\left| {{\psi _{\mathbf{k}}}\left( t \right)} \right\rangle  = {e^{ - iH_f\left( {\mathbf{k}} \right)t}}\left| {{\psi _{\mathbf{k}}}\left( {t = 0} \right)} \right\rangle  \hfill \\
   =& \left( \begin{gathered}
   - i\sin \left( {\tilde t/2} \right)\left( {{{\hat h}_x} - i{{\hat h}_y}} \right) \hfill \\
  \cos \left( {\tilde t/2} \right) + i\sin \left( {\tilde t/2} \right){  {{\hat h}_z}} \hfill \\
\end{gathered}  \right)\equiv\left| {{\psi _{\mathbf{k}}}\left( \tilde t \right)} \right\rangle, \hfill \\
\end{aligned}
\end{equation}
where the evolution time is rescaled as
\begin{equation}
  \tilde t ({\bf{k}}) = 2 h_f({\bf{k}}) t.  
\end{equation}
As mentioned in Sec.~\ref{Sec:Link_QAH}, experimentally, the rescaling factor $h_f({\bf{k}})$ can be measured using the spin-injection spectroscopy technique~\cite{Wang-Zhang2012,Cheuk-Zwierlein2012}. 
The Bloch vector for each quasimomentum and the rescaled time is then given by
\begin{equation}
  {\bf{n}}\left( {{\bf{k}},\tilde t} \right) =  - \left\langle {{\psi _{\bf{k}}}(\tilde t)} \right|\bm{\sigma} \left| {{\psi _{\bf{k}}}(\tilde t)} \right\rangle ,
\end{equation}
where the three components of ${\bf{n}}$ are as follows:
\begin{equation} \label{Eq:n_x_y_z}
  \begin{array}{l}
{n_x}\left( {{\bf{k}},\tilde t} \right) = \sin (\tilde t){{\hat h}_y} + 2{\sin ^2}(\tilde t/2){{\hat h}_x}{{\hat h}_z},\\
{n_y}\left( {{\bf{k}},\tilde t} \right) =  - \sin (\tilde t){{\hat h}_x} + 2{\sin ^2}(\tilde t/2){{\hat h}_y}{{\hat h}_z},\\
{n_z}\left( {{\bf{k}},\tilde t} \right) = \cos (\tilde t) + 2{\sin ^2}(\tilde t/2)\hat h_z^2.
\end{array}
\end{equation}
The typical results of the Bloch vector ${\bf{n}}\left( {{\bf{k}},\tilde t} \right)$ are shown in Fig.~\ref{Fig5}. We see that the Bloch vector is periodic in both the quasimomentum and the rescaled time. As such, the parameter space of $({ k_x,k_y,\tilde t})$ is a three-dimensional torus $T^3 = S_1\times S_1\times S_1$, i.e., a tensor product of three one-dimensional spheres $S_1$. 

The Berry connection in terms of the Bloch state $\left| {{\psi _{\mathbf{k}}}\left( \tilde t \right)} \right\rangle$ is defined by~\cite{Wilczek_Zee1983}:
\begin{equation}\label{Eq:A_mu}
  {A_\mu }\left( {{\bf{k}},\tilde t} \right) = \frac{i}{{2\pi }}\left\langle {{\psi _{\bf{k}}}(\tilde t)} \right|{\partial _\mu }\left| {{\psi _{\bf{k}}}(\tilde t)} \right\rangle ,
\end{equation}
where the index $\mu$ takes value in $k_x$, $k_y$, and $\tilde t$. The associated Berry curvature takes the form
\begin{equation}\label{Eq:J_mu}
  {J_\mu }\left( {{\bf{k}},\tilde t} \right) = {\epsilon _{\mu \nu \lambda }}{\partial _v}{A_\lambda }\left( {{\bf{k}},\tilde t} \right),
\end{equation}
where ${\epsilon _{\mu \nu \lambda }}$ is the total-antisymmetric tensor, and the Einstein's summation convention is used throughout. Importantly, while the Berry connection ${A_\mu }$ is gauge dependent, the Berry curvature $J_\mu$ is not. Indeed, $J_\mu$ is related to the experimental observable ${\bf{n}}\left( {{\bf{k}},\tilde t} \right)$ as~\cite{Wilczek_Zee1983,Wang2017}
\begin{equation}\label{Eq:J_mu_n}
  {J_\mu } = \frac{1}{{8\pi }}{\epsilon _{\mu \nu \lambda }}{\bf{n}} \cdot \left( {{\partial _\nu }{\bf{n}} \times {\partial _\lambda }{\bf{n}}} \right).
\end{equation}

From the Berry connection and the Berry curvature, the \emph{dynamical} Hopf invariant for the quenched system can be defined as~\cite{Wilczek_Zee1983}
\begin{equation}\label{Eq:Hopf_inv}
  \begin{aligned}
{I_H} &= \int {d{k_x}d{k_y}d\tilde t} {A_\mu }{J_\mu }\\
 &= \int {d{k_x}d{k_y}d\tilde t} {\epsilon _{\mu \nu \lambda }}{A_\mu }{\partial _v}{A_\lambda }.
\end{aligned}
\end{equation}
Here the integrand is a Chern-Simons density
\begin{equation} \label{Eq:rho_CS}
  {\rho _{{\rm{CS}}}} = {\epsilon _{\mu \nu \lambda }}{A_\mu }{\partial _v}{A_\lambda },
\end{equation}
which is apparently gauge dependent. Following from Eqs. (\ref{Eq:psi_t}), (\ref{Eq:A_mu}), and (\ref{Eq:J_mu}), a straightforward but cumbersome calculation yields the time-integrated Chern-Simons density as~\footnote{As the Chern-Simons density is gauge dependent, the following relationship is only valid for the specific gauge chosen in Eq.~(\ref{Eq:psi_t}), i.e., the Dirac string is located at the south pole.}
\begin{equation}
  {\bar \rho}_{\rm{CS}} \equiv  \int_0^{2\pi } {d\tilde t} {\rho _{{\rm{CS}}}} = \frac{1}{{4\pi }}{\bf{\hat h}} \cdot \left( {{\partial _{{k_x}}}{\bf{\hat h}} \times {\partial _{{k_y}}}{\bf{\hat h}}} \right).
\end{equation}
A further integration in the first Brillouin zone for both sides of the above formula gives [cf. Eqs.~(\ref{Eq:C_1_Qi}) and (\ref{Eq:Hopf_inv})]
\begin{equation}\label{Eq:I_and_C}
  I_H = C_1.
\end{equation}
Thus we have proved the dynamical Hopf invariant equals the Chern number of the post-quench Hamiltonian. 

We notice that Eq.~(\ref{Eq:I_and_C}) is proved within a different approach in Ref.~\cite{Wang2017}: There, one basically starts with Eq.~(\ref{Eq:J_mu_n}) as the definition of $J_\mu$, then solves $A_\mu$ according to Eq.~(\ref{Eq:J_mu}) for a particular gauge, and subsequently calculates the Hopf invariant in Eq.~(\ref{Eq:Hopf_inv}).

\subsection{Chern-Simons density through Bloch state tomography}
\begin{figure*}[htbp]
  \includegraphics[width=1.0\textwidth]{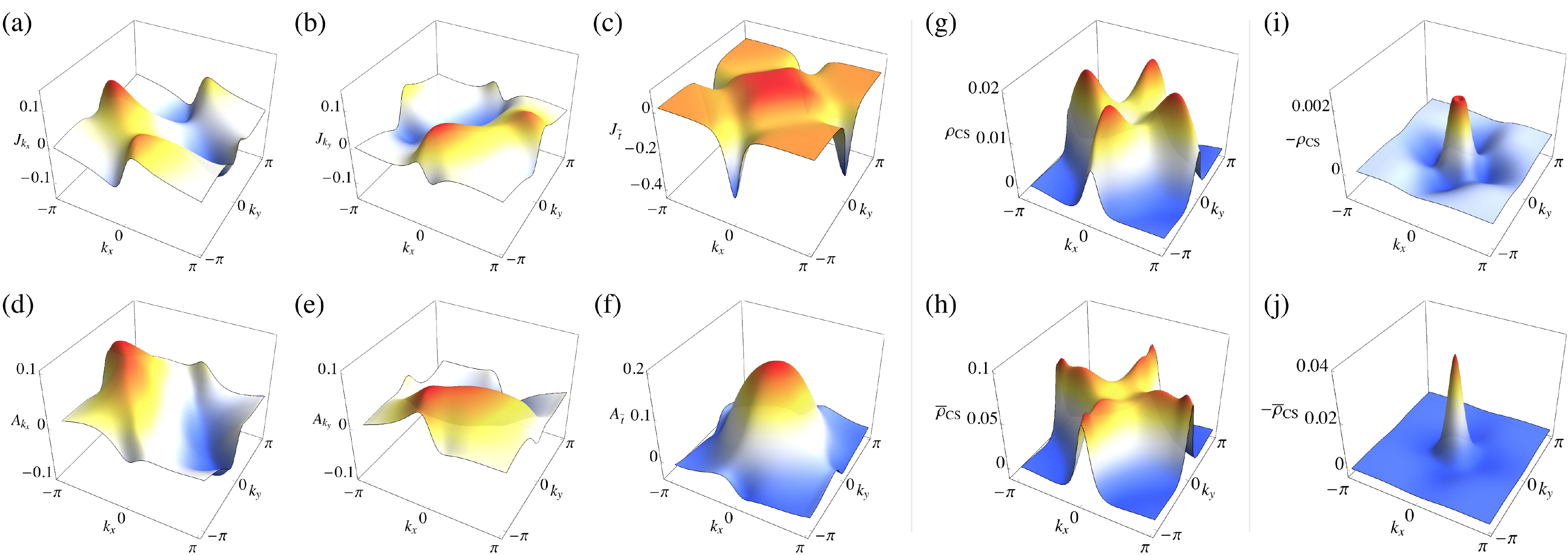}\\
  \caption{Determining the Hopf invariant for the quenched QAH model. (a-f) The Berry curvature $J_\mu$ ($\mu = k_x, k_y, {\tilde t}$) (a-c) and the corresponding Berry connection $A_\mu$ (d-f) in the gauge ${\partial _\mu }{A_\mu } = 0$, for a topological quench ($M_i = +\infty\to M_f = 0.6$) at rescaled time ${\tilde t}=0.4\pi$. (g) The Chern-Simons density ${\rho _{{\rm{CS}}}}=A_\mu J_\mu$ at ${\tilde t}=0.4\pi$ determined by the quantities shown in (a-f). (h) The time-integrated Chern-Simons density ${\bar \rho}_{\rm{CS}}$ for the same quench parameters as in (a-g). A further integration of ${\bar \rho}_{\rm{CS}}$ in the first Brillouin zone gives the Hopf invariant $I_H = 1$. (i-j) The Chern-Simons density ${\rho _{{\rm{CS}}}}$ at ${\tilde t}=0.4\pi$ (i) and the time-integrated Chern-Simons density ${\bar \rho}_{\rm{CS}}$ (j) for the case of a topologically trivial quench ($M_i = +\infty\to M_f = 2.5$). The quasi-momentum integration of ${\bar \rho}_{\rm{CS}}$ in the first Brillouin zone gives a trivial Hopf invariant $I_H = 0$ for this case. }\label{Fig6}
\end{figure*}

As is shown in Eq.~(\ref{Eq:J_mu_n}), the Berry curvature $J_\mu$ depends locally on the observable ${\bf{n}}\left( {{\bf{k}},\tilde t} \right)$. If we can find out a particular gauge to express the Berry connection $A_\mu$ as a function of the Bloch vectors ${\bf{n}}\left( {{\bf{k}},\tilde t} \right)$, or equivalently as a function of $J_\mu\left( {{\bf{k}},\tilde t} \right)$, then the Chern-Simons density and the dynamical Hopf invariant can be evaluated. To our limited knowledge, the expression of $A_\mu$ in Eq.~(\ref{Eq:A_mu}) with Eq.~(\ref{Eq:psi_t}) could not fulfill this task. As an alternative, we can take the following gauge
\begin{equation}
  {\partial _\mu }{A_\mu } = 0,
\end{equation}
and then solve Eq.~(\ref{Eq:J_mu}) to give~\cite{Wilczek_Zee1983}
\begin{equation}
  {\partial ^{ 2}}{A_\mu } =  - {\varepsilon _{\mu \nu \lambda }}{\partial _v}{J_\lambda }.
\end{equation}
Using the point-source Green's function technique, we get $A_\mu$ as a function of $J_\mu$ as follows~\cite{Jackson1999}:
\begin{equation}\label{Eq:A_as_J}
  \vec A(\vec K) = \frac{1}{{4\pi }}\int {d{{k}_x^\prime}d{{k}_y^\prime}d\tilde t^\prime} \vec J(\vec K') \times \frac{{\left( {\vec K - \vec K'} \right)}}{{|\vec K - \vec K'{|^3}}},
\end{equation}
where we denote $\vec K = (k_x, k_y, {\tilde t})$, $\vec A=(A_{k_x}, A_{k_y}, A_{\tilde t})$, and $\vec J=(J_{k_x}, J_{k_y}, J_{\tilde t})$. 
Thus we see that, in the gauge ${\partial _\mu }{A_\mu } = 0$, the Berry connection $A_\mu$ can be expressed as a (nonlocal) function of the Berry curvature $J_\mu$. 
This way, the Chern-Simons density ${\rho _{{\rm{CS}}}} = A_\mu J_\mu$ can be evaluated directly from the experimentally measured Bloch vectors ${\bf{n}}\left( {{\bf{k}},\tilde t} \right)$ in terms of Eqs.~(\ref{Eq:J_mu_n}) and (\ref{Eq:A_as_J}). The typical results of $A_\mu$, $J_\mu$, and $\rho_{\rm{CS}}$ are shown in Fig.~\ref{Fig6}. 

\section{Conclusion and outlook}\label{Sec:Conclusion}
Summarizing, motivated by the experimental realization of the QAH model in a topological Raman lattice by the PKU-USTC group~\cite{USTC_SOC2016}, we have discussed the quench dynamics of this system. We show how the Hopf link between trajectories with maximal spin imbalance can be determined by the spin-resolved time-of-flight measurement, and how the Hopf links for more than two preimages can be measured using additional rf pulses. Such additional rf manipulations combined with spin-resolved time-of-flight measurement accomplish quasimomentum- and time-resolved Bloch state tomography. In addition to the Hopf links, the Hopf invariant can be extracted directly. In a very recent experimental work by the PKU-USTC group, the Hopf link between preimages of $n_z=\pm 1$ for the QAH model has been measured~\cite{USTC_2018}. The remaining parts discussed in this work are also experimentally feasible with state-of-the-art techniques. 

In the current work, we have confined our discussion on the quench dynamics of the QAH model starting from a topologically trivial initial state. The quench dynamics with a topological initial state will be explored in the future~\cite{Xin2018}.

\begin{acknowledgments}
The author would like to thank Shuai Chen, Xin Chen, Xiong-Jun Liu, Chang-Rui Yi, and Hui Zhai for valuable discussions, and Ying Hu for carefully reading the manuscript.
This work is supported by MOST under Grant No. 2016YFA0301600 and NSFC Grant No. 11734010. 
\end{acknowledgments}

%

\end{document}